# Single-atom-resolved vibrational spectroscopy of a dislocation

Hailing Jiang[1], Tao Wang[1,2]*, Zhenyu Zhang[1], Ruochen Shi[2,3], Xifan Xu[1], Bowen Sheng[1], Fang Liu[1], Weikun Ge[1], Ping Wang[1], Bo Shen[1,3], Peng Gao[2,3], Lucas R Lindsay[4]*, and Xinqiang Wang[1,3]*

[1]State Key Laboratory for Mesoscopic Physics and Frontiers Science Center for Nano-optoelectronics, School of Physics, Peking University, Beijing 100871, China.

[2]Electron Microscopy Laboratory, School of Physics, Peking University, Beijing 100871, China.

[3]International Center for Quantum Materials, School of Physics, Peking University, Beijing 100871, China

[4]Materials Science and Technology Division, Oak Ridge National Laboratory, Oak Ridge, TN 37831, USA

*Corresponding author. E-mail: cwwangtao@pku.edu.cn, lindsaylr@ornl.gov, wangshi@pku.edu.cn

This manuscript has been authored by UT-Battelle, LLC under Contract No. DE-AC05-00OR22725 with the U.S. Department of Energy. The United States Government retains and the publisher, by accepting the article for publication, acknowledges that the United States Government retains a non-exclusive, paid-up, irrevocable, world-wide license to publish or reproduce the published form of this manuscript, or allow others to do so, for United States Government purposes. The Department of Energy will provide public access to these results of federally sponsored research in accordance with the DOE Public Access Plan (http://energy.gov/downloads/doe-public-access-plan).

**Phonon resistance from dislocation scattering is often divided into short-range core interactions and long-range strain field interactions. Using electron energy-loss spectroscopy on a GaN dislocation, we report observations of vibrational modes localized at specific core atoms (short-range) and strain-driven phonon energy shifts around the dislocation (long-range). Ab initio calculations support these findings and draw out additional details. This study reveals atomically resolved vibrational spectra of dislocations, thus offering insights for engineering improved material functionalities.**

*Introduction.*—Crystal imperfections, such as dislocations, scatter phonons and enhance thermal resistance[1-14]. As dislocations are common in III-nitride semiconductors, clarification of phonon-dislocation scattering channels will inevitably lead to better thermal management for semiconductor devices. Current experimental investigations into phonon-dislocation interactions predominantly focus on averaged bulk material properties[8-10]. Such macroscopic measurements lack details regarding mode-resolved and atomic-scale interactions which are important for developing new insights into vibrational transport behaviors. Theoretically, the mechanism of phonon-dislocation scattering can be divided into two regimes[15]: *short-range interactions*[16] localized at dislocation cores and *long-range interactions*[17] caused by the elastic strain field surrounding the dislocations. Despite this relatively simple theoretical decomposition, there is a lack of experimental studies of microscopic phonon behaviors near dislocations. Such measurements are challenging, as probing the atomic vibrational modes of individual dislocation cores requires atomic spatial resolution. Traditional techniques, such as Raman scattering spectroscopy, provide average, long-wavelength information for optical phonons in semiconductors with specific dislocation densities[8-10] or macroscopic strains[2]. Consequently, experimental measurement of mode-resolved phonon behaviors at dislocations with atomic resolution is not available.

To address these challenges, we investigate vibrational behaviors of 8-atom ring edge dislocations in GaN as shown in the HAADF-STEM image of Fig. 1b. The configuration of the 8-atom ring structure is distinguished by an atomic column with an under-coordinated center exclusively with Ga-N bonds. This kind of dislocation is the most common in wurtzite III-nitride semiconductors and is characterized by a highly symmetric atomic arrangement and strain field[18,19]. Recent developments in instrumentation have made coupling of scanning transmission electron microscopy (STEM) and electron energy-loss spectroscopy (EELS), thus providing measurement ability with atomic-scale spatial resolution and sub-10 meV energy resolution, enabling characterization of vibrational signals in various materials[20-30], defects[31-36], and interfaces[35-42]. Using this approach, vibrational modes of specific atoms in dislocations can be directly detected, providing an experimental way to clarify how dislocations specifically affect phonon behaviors.

In this study, we report two primary localized phonon peaks ($P_1$ and $P_2$) in the vibrational spectra at the dislocation cores, with energies of 55 meV and 109 meV, respectively, important for understanding short-range phonon-dislocation interactions. Density functional theory (DFT) calculations show that $P_1$ and $P_2$ originate from the vibration of nitrogen atoms in the center of the dislocation core along $[10\bar{1}0]$ and $[11\bar{2}0]$ directions, respectively. Furthermore, this 8-atom ring edge dislocation leads to a symmetric strain distribution from 3% compressive to 3% tensile strains within a 1 nm range, resulting in a phonon peak shifts of ~3 meV and ~1 meV for optical and acoustic phonons, respectively, important for understanding long-range phonon-dislocation interactions. This work provides important atomically resolved structural and vibrational details of dislocations that underlie how these interact with phonons, providing insights into the dislocation-related physics in semiconductors.

*Experiment and calculation details.*—Fig. 1a presents a schematic of the STEM-EELS used in our experiments. EELS is performed with a Nion U-HERMES200 STEM equipped with monochromator and aberration correctors. An electron beam, accelerated by 60 kV, is directly focused on the sample to achieve high spatial resolution at the atomic level. A large beam convergence angle of 35 mrad was used (see Fig. S1). Additionally, an EELS aperture on the diffraction plane, with a collection angle of 25 mrad, selectively collects scattered electrons covering multiple Brillouin zones (BZs). The collected spectrum represents the local phonon density of states (DOS). Subsequently, a magnetic prism deflects the electrons, enabling the collection of vibrational signals of the sample based on the energy loss of the electron beam. In this configuration, a spatial resolution of 0.1 nm and an energy resolution of 8 meV can be achieved. Fig. 1b displays the high-angle annular dark-field (HAADF) STEM image of an 8-atom ring edge dislocation in GaN along the [0001] zone axis. The 8-atom ring edge dislocation contains exclusively Ga-N bonds that are under-coordinated in a column at the center core. The experimental area of EEL spectra is marked by a white dashed rectangle covering a region of 2×2 nm$^2$. Detailed structural images of the 8-atom ring edge dislocation are given in Fig. S2. Based on the atomic image in Fig. 1c, we built atomic models of the 8-atom ring edge dislocation in GaN using 120-atom orthorhombic unit cells that consisted of the 8-atom ring structure (16 atoms) surrounded by hexagonal ring structures (104 atoms). The finite size atomic structures are periodically replicated in the *z* direction to create columns that are surrounded by vacuum in the other two directions for DFT simulations of the vibrational spectra. The surface bonds are saturated by fractional hydrogen atoms [43] to eliminate spurious effects of dangling bonds. Please see supplemental materials for detailed experimental and theoretical information.

*Dislocation cores (short-range).*—First, we measured the localized phonon modes at the core of a GaN 8-atom ring edge dislocation. The HAADF-STEM image of the dislocation along the [0001] zone axis is shown in Fig. 2a, which has the same area as that in the EELS measurements. The atomic positions of the Ga and N atoms are marked by blue and white balls, respectively. The center nitrogen atom of the dislocation is labeled A and some neighboring nitrogen atoms are labeled B (nearest nitrogen neighbor) and C (further nitrogen neighbor). To clarify the position of atoms, the corresponding atomic structure diagram of Fig. 2a is shown in Fig. 2b. We can get a phonon intensity map in atomic resolution by integrating the EELS intensity in the energy range of the acoustic phonon peaks to eliminate the influence of polariton excitations. Fig. 2c shows the measured EELS map of the $E_2$ phonon mode in the energy range of 19-24 meV. It is shown that the atomic structure can be distinguished even on the phonon map. The outline of the 8-atom ring is marked by white dashed lines on the figure, which agrees with those shown in Fig. 2a and 2b, respectively. The unobstructed EELS map is presented in Fig. S3, from which the position of the 8-ring atom can be seen.

To investigate vibrational variations within dislocation cores, we compared nitrogen vibrational modes of atom A with those of atoms B and C on both the left and right sides of the core atoms. EEL spectra of atoms A, B, and C are shown in Fig. 2d, which are extracted from the areas of the red, green, and blue rectangles in Fig. 2c, respectively. Measured EEL spectra of bulk GaN is also shown for comparison. We averaged the EEL signals of the symmetric atoms on either side of the dislocation. The separate left and right side EEL spectra of atoms B and C are depicted in Fig. S4. There are three dominant EEL spectral peaks for bulk GaN, $E_2$, $B_1$, TO/LO, at energies of 23, 41, and 76 meV, respectively. The $E_2$ mode and the mixed TO/LO phonon modes are also seen in the EEL spectra of atoms A, B, and C. There is a slight red shift of ~4 meV of the $B_1$ mode in the spectra of atoms A, B, and C in comparison to those of bulk GaN, as highlighted by the red (A) and black (bulk GaN) dash line in Fig. 2d. In comparison with those of bulk GaN, the EEL spectra localized at atoms A, B, and C all reveal two new vibrational modes, labeled here as $P_1$ and $P_2$. Energies of the $P_1$ (55 meV) and $P_2$ (109 meV) modes are nearly the same in the spectra of all three nitrogen atoms. However, the intensity of the $P_1$ mode becomes weaker at B and C atoms compared with that for the A atom. On the contrary, intensity of the $P_2$ mode remains almost constant among these atoms.

Using the Quantum Espresso[44] package and the local density approximation (LDA) within density functional theory (DFT), we calculated the vibrational atom-resolved partial densities of states (pDOS) of atoms A, B, and C in the 8-atom ring edge dislocation and compare with bulk GaN in Fig. 2e. We note that our calculated results for bulk wurtzite GaN[45] show excellent agreement with phonon frequencies observed from inelastic X-ray scattering[46]. The three dominant EEL spectral peaks detected from bulk GaN agree with the calculations. The redshift of the $B_1$ mode is also verified in the calculations shown in Fig. 2e, difference between marked red and black dashed lines. The calculations of the phonon spectrum of atom A exhibits $P_1$ and $P_2$ modes at the energies of 48.8 and 92.5 meV, respectively. The slight discrepancy from the experimental results may result from the limited energy resolution of EELS and the choice of pseudopotentials and DFT flavor in the calculations. The peak intensities of the $P_1$ and $P_2$ modes are much weaker in the calculated phonon spectra than those experimentally detected for atoms B and C. Only a small peak indicating the $P_2$ mode is discernible in the calculated phonon spectrum of atom B.

To clarify the nature of the $P_1$ and $P_2$ modes, the calculated and experimental localized phonon intensity maps were extracted at their corresponding energies, as shown in Fig. 2f. The atomic structure of the 8-atom ring along with calculated atomic vibrational eigenvectors of $P_1$ and $P_2$ modes are marked on the EELS maps. It is shown that the $P_1$ mode mainly originates from the vibrations of the N atoms at the position of A and the atom below A along the $[10\bar{1}0]$ direction. The $P_2$ mode originates from the vibration of N atoms at positions A and B along the $[11\bar{2}0]$ direction. This corresponds to the intensity distribution of experimental and calculated EELS maps in the energy range of $P_1$ and $P_2$, demonstrating agreement between calculation and measurement. Fig. S5 illustrates the dislocation component obtained by subtracting the spectra of surrounding atoms C from those of the central atoms A and B, in both calculations and experiments. It shows the phonon intensity reductions of $P_1$ and $P_2$ going from the C atoms to A and B atoms.

*Dislocation strains (long-range).*—After investigating the localized phonon modes caused by dislocation cores, we examined the influence of the induced strain field around the dislocation on phonons. Geometrical phase analysis (GPA) was performed on the HAADF-STEM image within the white dashed square in Fig. 1b. The strain distribution, $\varepsilon_{xx}$, for the 8-atom ring edge dislocation is presented in Fig. 3a, revealing compressive and tensile strains of about 3% (equivalent to pressure of about 8 GPa) above and below the dislocation in Fig. 1b, respectively. The peak mapping of optical phonons (Fig. 3b) demonstrates an energy increase above and a decrease below the 8-atom ring in the figure. This corresponds to the GPA map in Fig. 3a, which shows compressive strain above and tensile strain below the 8-atom ring, confirming the phonon shift due to strain. To further quantify energy shift with strain, STEM-EEL spectra (Fig. 3c) were extracted from the area of red and blue shaded rectangles in Fig. 3a. In the EEL spectra, under this approximate 6% strain difference, phonons exhibit an energy shift of approximately 3 meV (0.188 meV·GPa$^{-1}$) for optical phonons and 1 meV (0.063 meV·GPa$^{-1}$) for acoustic phonons. The weak intensities of P$_1$ and P$_2$ can also be detected in these spectra, which are influenced by the dislocation cores. The calculated pDOSs of GaN with compressive and tensile strain are shown in Fig. 3d, extracted from the atoms above and below the 8-atom ring depicted in Fig. S6. The overall features demonstrate agreement between experiments and calculated results. The minor variance between experimental outcomes and theoretical predictions may arise from limited energy resolution in EELS measurements. To further confirm the shift of acoustic phonons, EELS intensity maps of the 8-atom ring edge dislocation are shown in Fig. S7 in the energy ranges of 19-21 meV and 24-26 meV. This provides a clear visualization of the atomic structure of the 8-atom ring edge dislocation (marked by the red circle). Remarkably, an increase in phonon intensity is observed in the compressively strained upper region at 24-26 meV and in the tensile strained lower region at 19-21 meV. This observation reaffirms that acoustic phonon energies increase with compressive strain and decrease with tensile strain. The red shift in phonon energies under tensile strain can be attributed to increased atomic spacing, which diminishes the interaction forces and consequently reduces phonon energies. Conversely, compressive strain induces an opposite effect.

*Conclusions.*—In summary, the single-atom vibrational spectroscopy of 8-atom ring edge dislocation in GaN was visualized through STEM-EELS. The phonons scattered by center core atoms result in two local vibrational modes, P$_1$ and P$_2$, caused by the vibrations of the center nitrogen atoms along the [10$\bar{1}$0] and [11$\bar{2}$0] directions, respectively. Additionally, the tensile strain causes a red shift in phonon energy, while the compressive strain causes a blue shift. Overall, our study provides experimental observations of both short-range and long-range atomic-level vibrational behaviors of dislocations, important to understanding phonon-dislocation interactions. These findings offer new insights into dislocation-driven phonon dynamics and provides key details regarding common limitations in device performance.

This work was supported by the the National Key R&D Program of China (No. 2023YFA1407000), the National Natural Science Foundation of China (No. 62321004, 62104010, 62227817, and 62374010), and the Beijing Natural Science Foundation (No. Z200004). This work was supported by the High-performance Computing Platform of Peking University. We acknowledge the Electron Microscopy Laboratory of Peking University for the use of electron microscopes. Calculations and manuscript development (L.L.) were supported by the U.S. Department of Energy, Office of Science, Office of Basic Energy Sciences, Material Sciences and Engineering Division. The calculations used resources of the Compute and Data Environment for Science (CADES) at the Oak Ridge National Laboratory, which is supported by the Office of Science of the U.S. Department of Energy under Contract No. DE-AC05-00OR22725, and resources of the National Energy Research Scientific Computing Center, which is supported by the Office of Science of the U.S. Department of Energy under Contract No. DE-AC02-05CH11231.

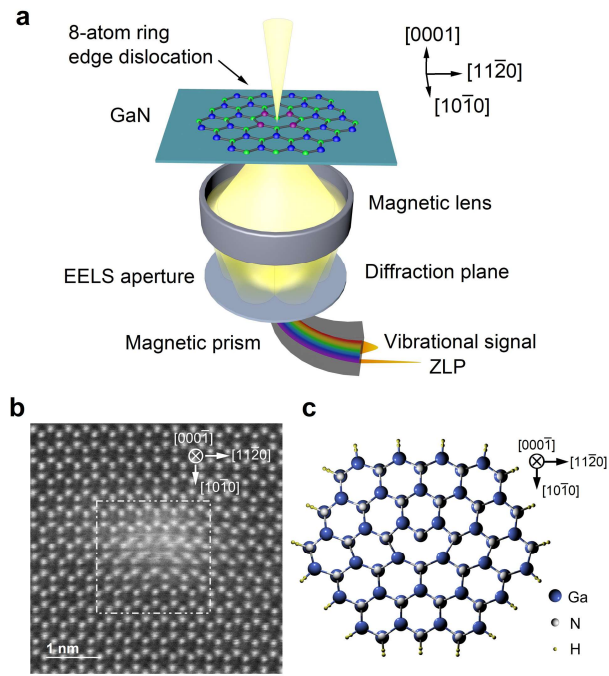

**Fig. 1 | Experimental setup and 8-atom ring edge dislocation structure. a.** Schematic of electron energy-loss spectroscopy incorporated in a scanning transmission electron microscope. **b.** HAADF-STEM image of a GaN 8-atom ring edge dislocation viewed from the [0001] zone axis. The white dashed rectangle marks the area where the EELS experiment was performed. **c.** The calculated atomic structure of the 8-atom ring edge dislocation. Blue, white and yellow balls are Ga, N and H atoms, respectively.

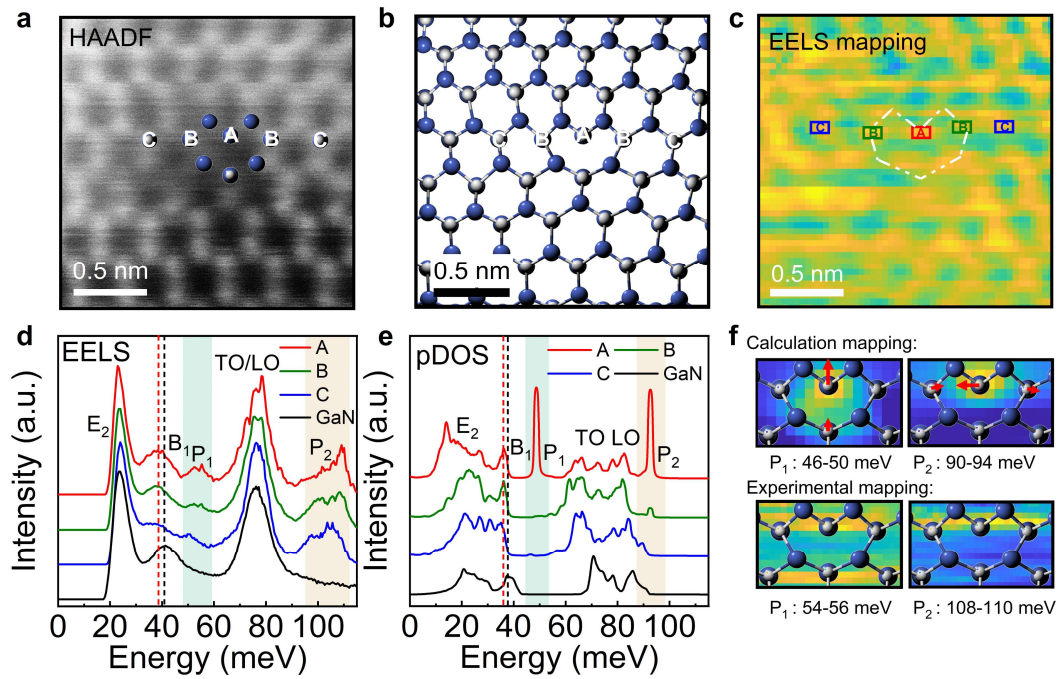

**Fig. 2 | The atomic-resolution phonon measurement of atoms in an 8-atom ring edge dislocation. a.** HAADF-STEM image of GaN with an 8-atom ring edge dislocation projected along the [0001] zone axis. The central atom (A) of the dislocation and its neighboring atoms (B and C) are marked. **b.** Atomic structure corresponding to the same region as **a**. **c.** EELS intensity maps formed by signal integration in the energy range of 19-24 meV, corresponding to the same region as **a**. The outline of the 8-atom ring is marked by white dashed lines in the figure. **d.** EEL spectra extracted from the rectangles A (red), B (green), and C (blue) in **c** and bulk GaN(black). Bulk optical and acoustic phonon peaks as well as localized phonon modes $P_1$ (cyan) and $P_2$ (brown) are marked on the spectra. The red and black vertical dotted line shows the phonon energy of the $B_1$ mode in atom A and bulk GaN, respectively. **e.** Calculated pDOS of the 8-atom ring edge dislocation corresponding to atoms A (red), B (green), C (blue), and bulk GaN (black). The bulk phonon peaks and localized modes $P_1$ (cyan) and $P_2$ (brown) are shown in the figure. **f.** Phonon eigenvectors for localized phonon modes $P_1$ and $P_2$ along with the localized EELS maps at the corresponding energy. The red arrows show the directions and vibration amplitudes of the eigenvectors.

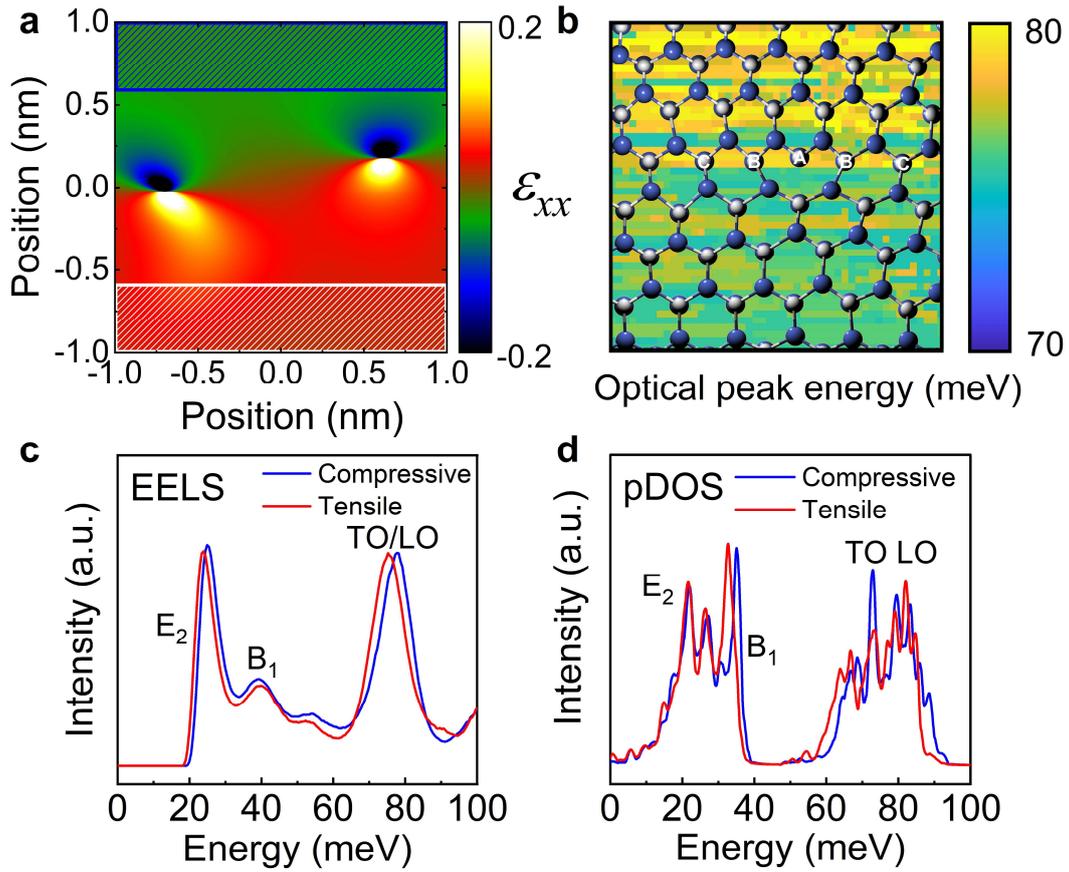

**Fig. 3 | The defect phonon affected by strain**. **a.** GPA map corresponding to the image in the white dashed rectangle of Fig. 1c showing the lattice strain of components $\varepsilon_{xx}$. **b.** The peak mapping of optical phonons corresponds to the same special region in **a**, with the atomic structure overlaid. As shown, the phonon energy of atoms above the 8-atom ring under compressive strain is higher than that of atoms below the 8-atom ring under tensile strain. The overall features of the peak mapping are similar to the GPA map in **a**. **c.** Phonon spectra with compressive strain (blue) and tensile strain (red) in measurements, which were integrated from the blue shaded regions and white shaded regions in Fig. 3a, respectively. **d.** The simulated phonon spectra of compressive strain (blue) and tensile strain (red) extracted from the atoms above and below the 8-atom ring in Fig. S7.

## Supplemental Materials

## Methods

**Sample preparation and characterization.** Wurtzite quasi-vdW epitaxial GaN films were grown on graphene/sapphire substrates using plasma-assisted molecular beam epitaxy (MBE). As the growth progressed, the nucleation islands coalesced and eventually formed a GaN film with 8-atom ring dislocations. A transmission electron microscopy (TEM) lamella was cut by a Thermo Fisher Scientific Helios G4 UX focused ion beam (FIB) system. To determine the atomic arrangement of the GaN planar defect, scanning transmission electron microscopy (STEM) measurements were performed using a spherical aberration-corrected FEI Titan Cubed Themis G2 300.

**Electron energy loss spectroscopy (EELS) data acquisition.** STEM-EELS measurements were performed using a Nion HERMES 200 microscope operating at 60 kV at room temperature equipped with both monochromator and aberration correctors. EELS datasets were obtained with a convergence semiangle of 35 mrad and a collection semiangle of 25 mrad. The STEM-EELS on-axis dataset was acquired in a 50×50 pixel mapping area covering a region of $2 \times 2$ nm$^2$ that included the 8-atom ring dislocation. The dwell time for each pixel was 200 ms, resulting in a total acquisition time of approximately 8 minutes for each dataset. The sample drift during the acquisition was typically less than 0.1 nm. The energy resolution achieved under these conditions was approximately 10 meV, while the spatial resolution was approximately 0.1 nm.

*Ab initio* **calculations.** All density functional theory (DFT) calculations were done using the Quantum Espresso [1] package within the local density approximation (LDA) using norm conserving pseudopotentials (pz-bhs for Ga atoms and pz-vbc for N and H atoms). Bulk GaN calculations followed the procedures outlined in Ref. 2 for the phonons with a 1% lattice expansion (set 2 of that paper), which gives good agreement with measured data. The QE .scf input file and harmonic interatomic force constants (IFCs) are also given in the supplemental information.

Starting with the bulk GaN calculation and based on the atomic image in Fig. 1c of the manuscript, we built atomic models of the 8-atom ring edge dislocation in GaN by hand using 120-atom orthorhombic unit cells that consisted of the 8-atom ring structure (16 atoms) surrounded by hexagonal ring structures (104 atoms), see Fig. S7 and corresponding .scf input file in the Supplemental Materials. The finite size atomic structures are periodically replicated in the *z* direction (*c*-axis) to create columns that are surrounded by vacuum in the other two directions, *i.e.*, the orthorhombic lattice parameters were: *a* = 28.58 Å, *b* = 25.40 Å, and *c* = 5.24 Å. The surface bonds are saturated by fractional hydrogen atoms (38 atoms)[3] to eliminate spurious effects of dangling bonds. We used a wave function energy cutoff of 43.9 Ry, an energy convergence threshold of 1e$^{-16}$ Ry, and a 1×1×5 Γ-centered sampling mesh for the electronic structure calculations. We used Γ-point only sampling for the density functional perturbation theory calculations of the harmonic IFCs. Input and IFC files can be found in the Supplemental Information. The atom-resolved density of states were determined on a relative dense integration mesh along the *c*-axis (48 points using Gaussian integration quadrature) and 0.75 meV smearing in the Lorentzian representing energy conservation.

For the EELS mapping simulation, the electron beam was modeled as a Gaussian beam. The scattering cross section between electrons and phonons were calculated. The scattering cross section is represented as[4-6]:

$$\frac{d^2\sigma}{d\omega d\Omega} \propto \sum_{mode\ \lambda} |F_\lambda(\mathbf{q})|^2 \left[ \frac{n+1}{\omega_\lambda(\mathbf{q})} \delta(\omega - \omega_\lambda(\mathbf{q})) + \frac{n}{\omega_\lambda(\mathbf{q})} \delta(\omega + \omega_\lambda(\mathbf{q})) \right] \quad [1]$$

where $\omega_\lambda(\mathbf{q})$ and *n* are the frequency and occupancy number of the *λ*th phonon mode with wavevector **q**. The Dirac delta function is represented by *δ(x)*. The coupling factor is given by:

$$F_\lambda(\mathbf{q}) \propto \frac{1}{q^2} \sum_{atom\ k} \frac{1}{\sqrt{M_k}} e^{-i\mathbf{q}\cdot\mathbf{r}_k} e^{-W_k(\mathbf{q})} Z_k(\mathbf{q}) [\mathbf{e}_\lambda(k,\mathbf{q}) \cdot \mathbf{q}] \quad [2]$$

This factor is determined by the mass $M_k$, the effective charge $Z_k(\mathbf{q})$, the real-space position $\mathbf{r}_k$, the Debye-Waller factor[7] $e^{-2W_k(\mathbf{q})}$ and the phonon eigenvector $\mathbf{e}_\lambda(k,\mathbf{q})$ of the *k*th atom in a unit cell. The effective charge $Z_k(\mathbf{q})$ was calculated as per literature[6], with atomic form factors constructed from parameters in the literature[8]. When the collection range of the EELS aperture is large enough to include multiple BZs, the summation of the squared coupling factor approximately gives the phonon

DOS.

**EELS Data Processing.** For each 3D-EELS dataset, the EEL spectra were initially aligned using normalized cross-correlation to correct for beam energy drift. Then, the spectra were normalized by the zero-loss peak (ZLP) total intensity. To eliminate Gaussian noise, a block-matching and 3D filtering algorithm were applied [9,10]. The EELS background was fitted using a Pearson function [11] and subtracted. To enhance the resolution and reduce the broadening caused by finite energy resolution, Lucy-Richardson deconvolution was employed. The statistical factors of the background-subtracted spectra were further corrected, which resulted from the different scattering probabilities of phonons with different frequencies [12], to make it more comparable to the real phonon DOS.

## Author Contribution

T.Wang and X.Q.Wang conceived and supervised the project; H.L.Jiang performed the STEM-EELS experiment and data analysis assisted by T.Wang, Z.Y.Zhang and R.C.Shi with the guidance of X.Q.Wang; L.R. Lindsay performed DFT calculations assisted by H.L.Jiang; R.C.Shi designed the toolbox for data processing; F.Liu grew the samples with the guidance of X.Q.Wang; T.Wang prepared the TEM sample; R.C.Shi, B.Shen and W.K.Ge helped the data interpretation. H.L.Jiang wrote the manuscript assisted by T.Wang, B.W.Sheng, P.Wang and P.Gao under the direction of L.R. Lindsay and X.Q.Wang. All the authors contributed to this work through useful discussion and/or comments to the manuscript.

# Supplemental Figures

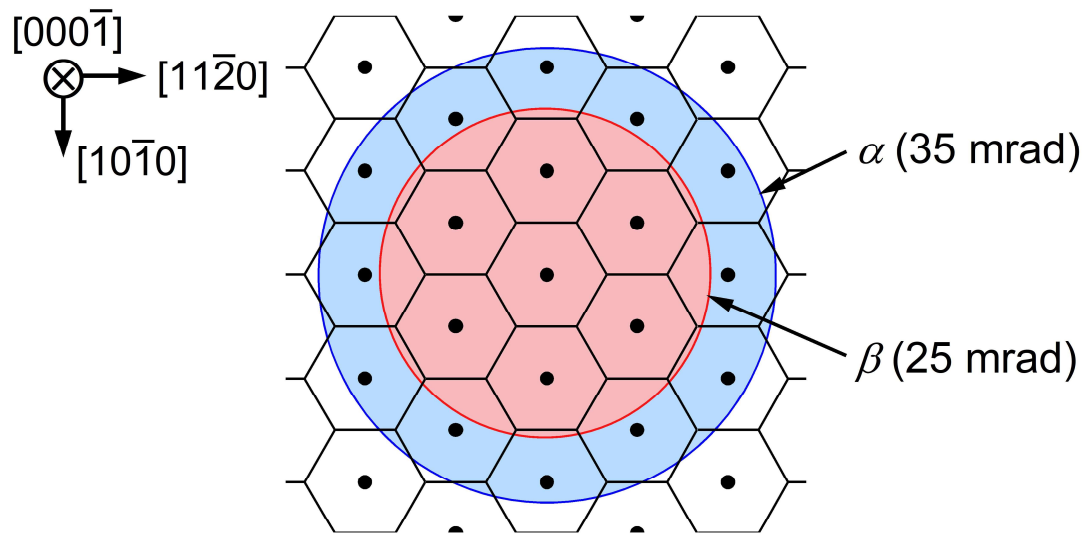

**Fig. S1. Schematic of the EELS aperture placement.** The convergence semi-angle $\alpha$ of the incident beam is 35 mrad, which corresponds to the bright field shown by the blue circle. The red circle marks the position of the EELS collection aperture, which covered an area of $\beta = 25$ mrad in the diffraction plane.

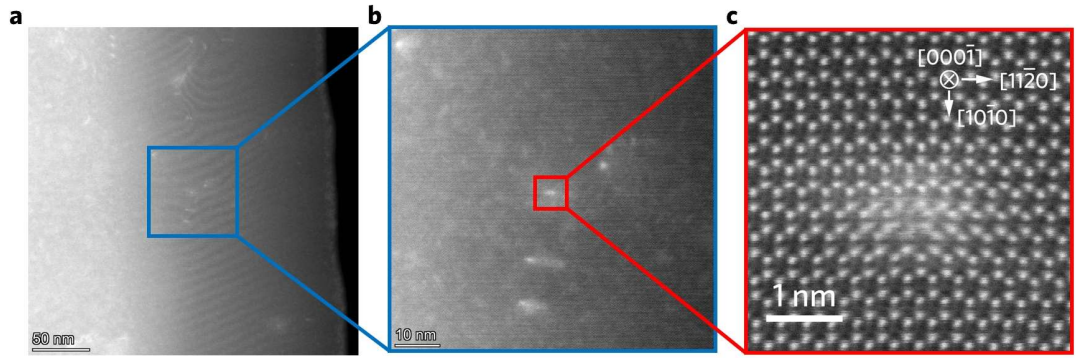

**Fig. S2. HAADF-STEM images around an 8-atom ring edge-component threading dislocation. a.** Large-scale HAADF-STEM image of the sample we used. The area contained the GaN 8-atom ring edge dislocation is marked by a blue rectangle. This image is viewed along the [0001] zone axis. The variation in atomic contrast within the HAADF-STEM signal serves as a key indicator of strain intensity and facilitates the location of dislocations. **b.** Zoom-in HAADF-STEM image of a GaN 8-atom ring edge dislocation obtained at the area marked by the blue rectangle in **a**. **c.** Zoom-in HAADF-STEM image of a GaN 8-atom ring edge dislocation obtained at the area marked by the red rectangle in **b**.

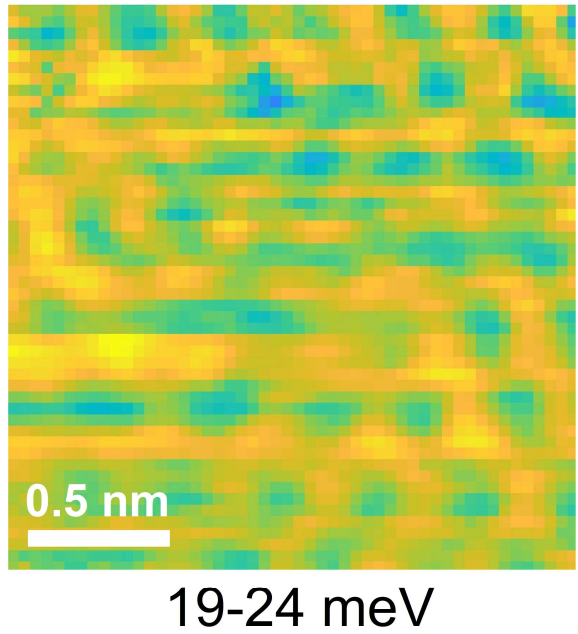

19-24 meV

**Fig. S3. Unobstructed EELS mapping**. Reproduction of Fig. 2c without obstructed atom annotation.

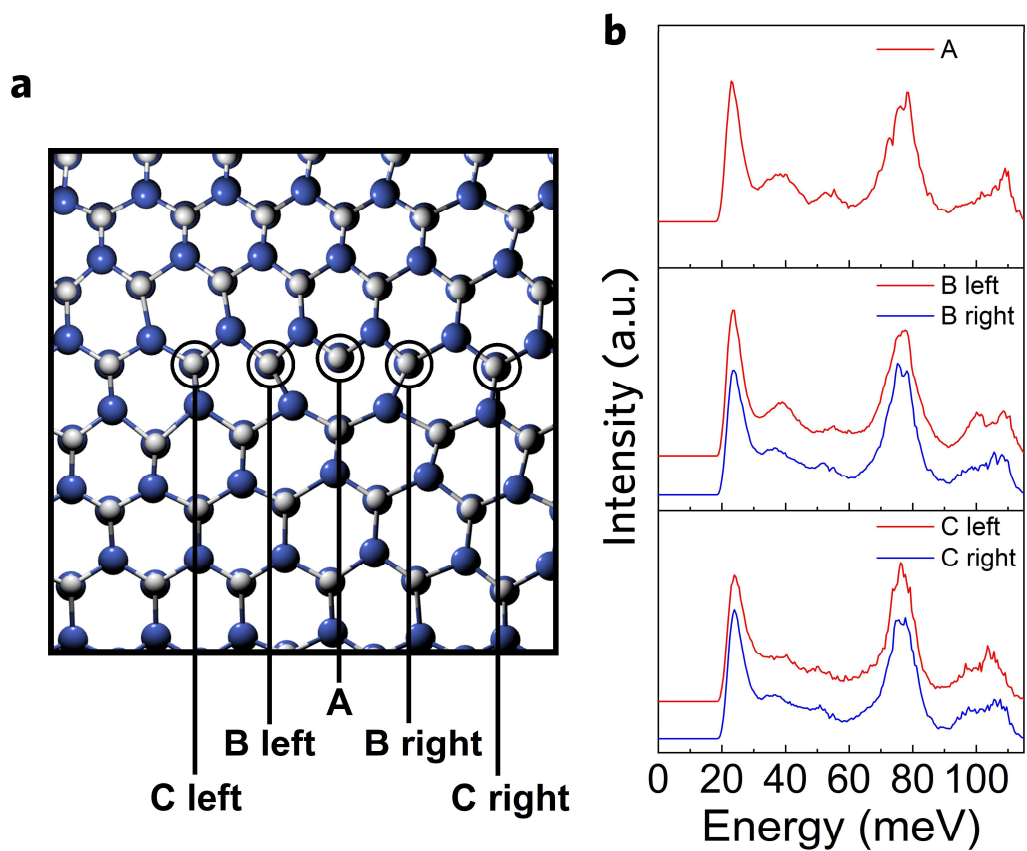

**Fig. S4. Phonon measurement of atoms on different sides of the dislocation**. **a.** Atomic structure of the spatial range in measurement. **b.** EELS spectra extracted from atoms on the left and right side in 8-atom ring dislocation, in contrast to Fig. 2d where the EELS signals of the B and C atoms on the left and right sides were averaged. The primary features of the vibrational spectra of B and C atoms on both sides are consistent. The atoms A, B left, B right, C left and C right are marked in the **a**.

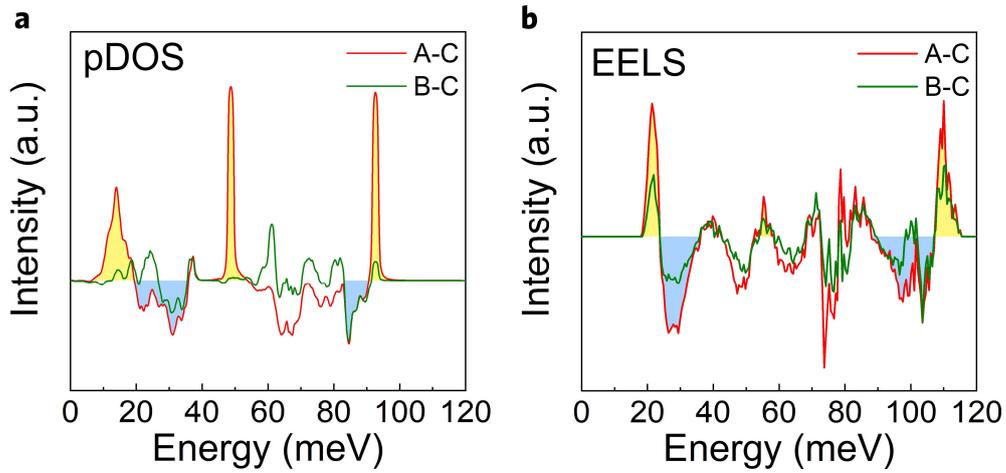

**Fig. S5. The difference of phonon spectra between atoms in 8-atom ring edge-component threading dislocation. a.** The green and red curves with gradient filling (where yellow and blue colors represent positive and negative residuals respectively) represent the dislocation component obtained by subtracting the calculated phonon pDOS of atoms C from that from atoms A (green curve) and atoms B (red curve) in Fig. 2e. **b.** The green and red curves with gradient filling (where yellow and blue colors represent positive and negative residuals, respectively) show the dislocation DOS components obtained by subtracting the EEL spectra of atoms C from atoms A (green curve) and from atoms B (red curve) in Fig. 2d. The difference in the experimental and theoretical EELS spectra of the core A atoms and the surrounding B and C atoms demonstrate spectral agreement with consistent regions of enhancement and reduction.

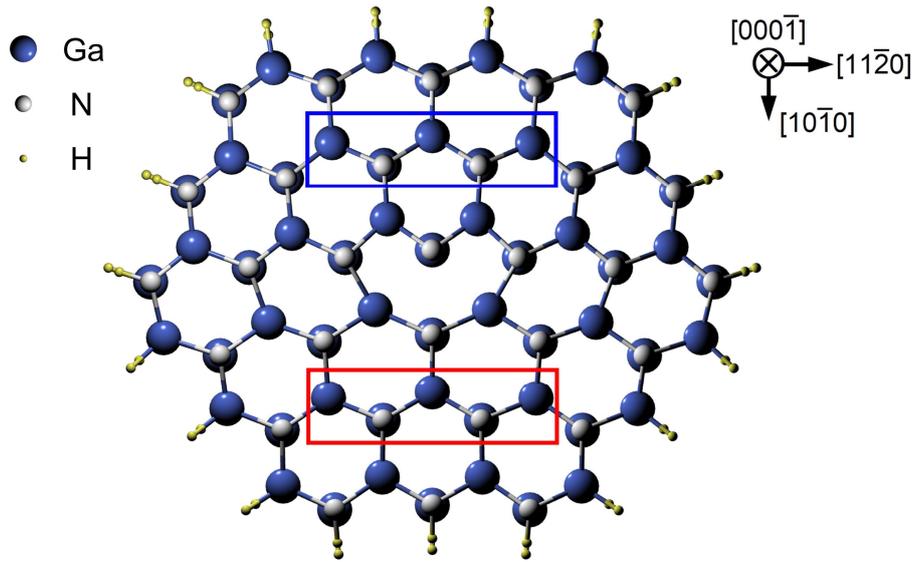

**Fig. S6. The range of calculated pDOSs of GaN with compressive and tensile strain.** The calculated phonons of GaN under compressive/tensile strain caused by the 8-atom ring dislocation are extracted from the averaged pDOSs of 10 atoms within the blue/red rectangle above/below the 8-atom ring.

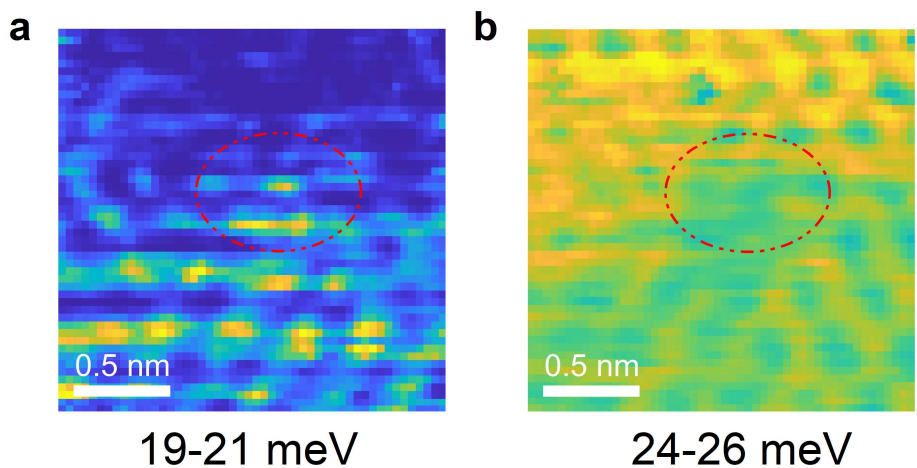

**Fig. S7. Intensity mapping of acoustic phonons in measurement. a.** EELS intensity maps formed by signal integration over the energy ranges of 19-21 meV, corresponding to the spatial region in Fig 2a. **b.** EELS intensity maps formed by signal integration over the energy ranges of 24-26 meV, corresponding to the spatial region in Fig 2a. The 8-atom ring is highlighted by a red circle. Notably, an increase in phonon intensity is observed in the compressively strained upper region within the higher energy range of 24-26 meV, while in the tensile strained lower region, the increase occurs within the lower energy range of 19-21 meV. These results confirm that acoustic phonon energies rise with compressive strain and decrease with tensile strain.